\documentclass[twocolumn,showpacs,preprintnumbers,amsmath,amssymb]{revtex4}
\usepackage{graphicx}
\usepackage{dcolumn}
\usepackage{bm}
\begin{document}

\preprint{}
\title{Fundamental characteristics of transverse deflecting fields}

\author{Valentin V. Paramonov} 
\email{paramono@inr.ru,}
\affiliation{Institute for Nuclear Research, 117312, Moscow, Russia}

\author{Klaus Floettmann}
\affiliation{DESY, Notkestr. 85, 22603 Hamburg, Germany}

\date{\today}
\begin{abstract}
The Panofsky-Wenzel theorem connects the transverse deflecting force in an rf structure with the existence 
of a longitudinal electric field component. In this paper it is shown that a transverse deflecting force
is always accompanied by an additional longitudinal magnetic field component which leads to an emittance growth 
in the direction perpendicular to the transverse force. Transverse deflecting waves can thus not be 
described by pure TM or TE modes, but require a linear combination of basis modes for their representation. 
The mode description is preferably performed in the HM--HE basis to avoid converge problems, which are 
fundamental for the TM--TE basis.  
\end{abstract}


\maketitle

\section{Introduction}Transverse deflecting rf fields find numerous applications in modern particle accelerators for example as 
particle separators~\cite{Edwards 2001} or fast rf deflectors~\cite{Alesini 2009}, as streaking device for 
diagnostics purposes~\cite{Ratner 2015}, in emittance exchange beam lines~\cite{Emma 2011} or as crab cavities 
in circular colliders as the LHC~\cite{Burt 2014}. It is well known, that a beam passing through a transverse 
deflecting field will not only receive the desired phase dependent transverse momentum, but it will also change 
its energy spread due to a longitudinal electric field component which varies over the transverse size of the beam. 
The fundamental relation of the transverse gradient of the longitudinal electric field component and the 
transverse momentum was formulated by Panofsky and Wenzel in their seminal paper in 1956~\cite{Pif}. 
Originally derived in the context of transverse deflecting rf structures, which is also the focus of this 
paper, the Panofsky-Wenzel theorem became a fundamental relation also for the discussion of wake 
potentials~\cite{Wilson} and devices as pickups and kickers~\cite{Lambertson}.\\    
Complementary to the longitudinal electric field component a longitudinal magnetic field component 
exists in transverse deflecting rf fields which has been widely ignored so far. The existence of the 
longitudinal magnetic component requires to revisit the general mode description of transverse deflecting 
rf structures. Due to the coupling of the transverse motion to the longitudinal magnetic field it leads 
to a small, but fundamental contribution to the transverse beam emittance in the direction perpendicular 
to the transverse force.
\section{Supporting structure}
To provide an effective interaction between a particle, moving with the velocity $v_z=\beta_z c$, and an 
electromagnetic rf field
it is necessary to match the phase velocity $v_{ph}$ of a harmonic field component to the velocity of 
the particle. Since the phase velocity in a simple waveguide is higher than the speed of light $c$, 
while $\beta_z \leq 1.0$ it is necessary to slow down the wave in an appropriate structure.\\
\begin{figure}[hhh]
    \centering
    \includegraphics*[width=85mm]{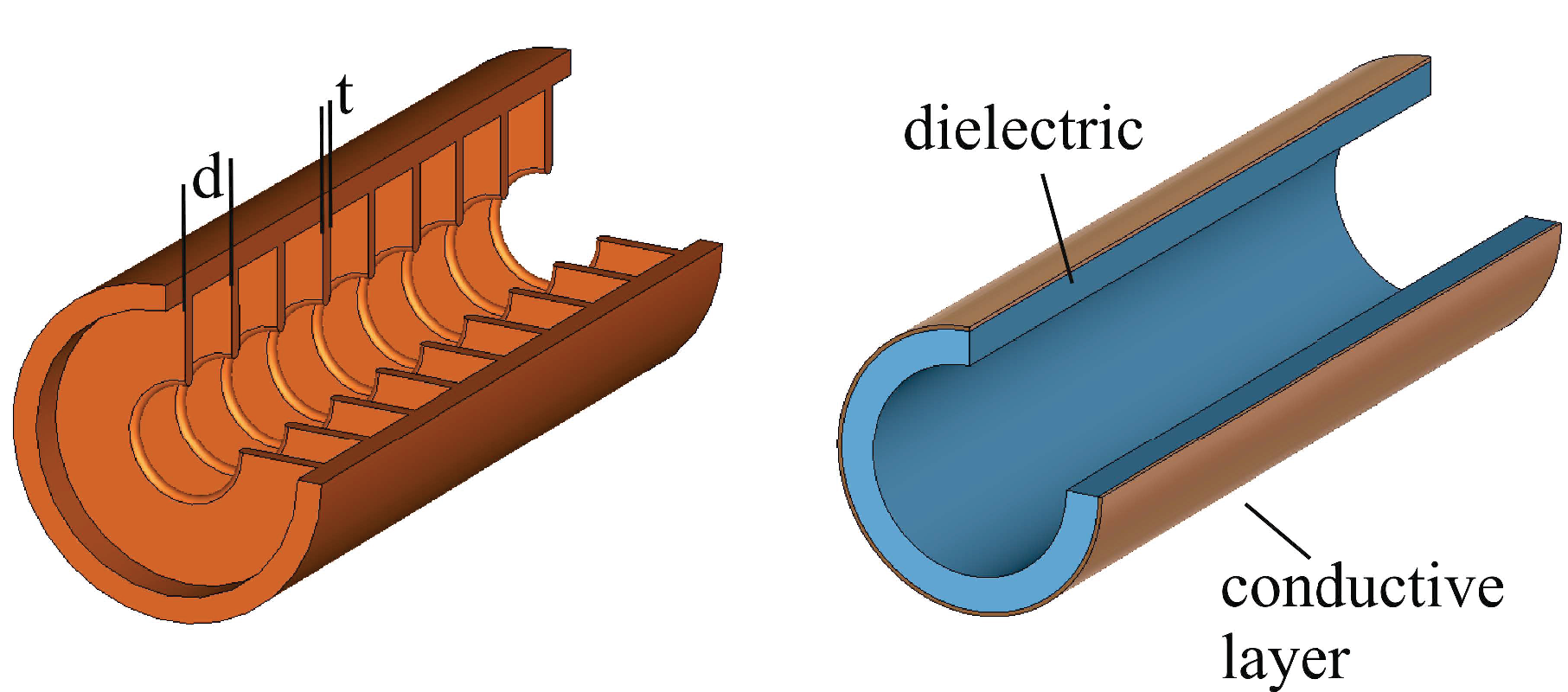}
    \caption{Supporting structures for slow deflecting waves: periodical iris loaded circular waveguide, 
left, and dielectric lined waveguide, right.}
    \label{fig_1}
\end{figure}
Commonly periodically iris loaded structures as the example shown in Fig.~\ref{fig_1}, left, are employed to achieve 
this. The field distribution in a periodical structure represents a sum of spatial harmonics and by proper 
selection of the period length $d= \frac{\beta_z \lambda \theta}{2 \pi}$, $0 \leq \theta \leq \pi$ a synchronous 
harmonic can be generated. $\theta$ is the phase advance per period and the wavelength $\lambda$ for this kind of 
structures is typically in the range of 20-3~cm (L-- to X--band). A fundamental property of 
periodical structures is the appearance of spatial harmonics which lead to nonlinearities in the field 
distribution. The nonlinearities can not be completely eliminated but they can be minimized in the region 
occupied by the beam by a proper design of the structure~\cite{FlPa2014, Paramonov2013}.\\
Spatial harmonics do not appear in structures which are uniform in the longitudinal $z$ direction. To slow down 
the wave the structure can be partially filled with a dielectric medium 
(see Fig.~\ref{fig_1}, right) but also thin metallic layers~\cite{Tsakanov} or even a rough surface~\cite{Novokhatsky} can 
lead to slow waves.\\
    Dielectric lined waveguides have been discussed already in the 1960s~\cite{Vagin, Dawson} as candidates for 
beam separators operating at typical rf frequencies (3~GHz). They gain now interest again as streaking device 
for diagnostics purposes operating in the sub-THz to THz range~\cite{dlw_tds}, where the dimensions are so 
small that the production of periodical structures reaches technical limits. The demand to resolve ever shorter 
bunch length and also the progress in the generation of THz pulses of sufficient power and pulse 
length~\cite{Ahr2017} are driving forces of these developments. Moreover, the absence of spatial harmonics makes 
dielectric lined waveguides also attractive from the beam dynamics point of view, because undesired side 
effects, like undesired emittance contributions, are minimized to their fundamental limits.
\section{General relations and mode description}  
The transverse deflecting force $\vec F_{\perp}$ acting on a particle moving along the longitudinal axis $z$ with 
the velocity $\vec V= \vec i_z v_z$ is defined by the transverse components of the Lorentz force
\begin{equation}
\begin{gathered}
\vec F_{\perp} =e(E_x-v_z B_y) \vec i_x+e(E_y+v_z B_x) \vec i_y,  \hfill \\
\vec F_{\perp} =e(E_r-v_z B_{\vartheta}) \vec i_r+e(E_{\vartheta}+v_z B_r) \vec i_{\vartheta},  \hfill
\end{gathered}
\label{Eq. 0}
\end{equation}
where $\vec i_x, \vec i_y, \vec i_z$ and $\vec i_{\vartheta}, \vec i_r, \vec i_z$ are the basis vectors of the Cartesian 
and cylindrical coordinate system, respectively.\\
The fundamental relations of deflecting field components and the corresponding longitudinal electric and magnetic
field components follow directly from Maxwell's equations 
$\text{curl} \vec E =  - \frac{\partial }{{\partial t}} \vec B$ and 
$\text{curl} \vec B =  \frac{\partial }{{\partial t}} \frac{\vec E}{c^2}$ as:
\begin{equation}
\begin{gathered}
\frac{\partial}{\partial x} E_z =\frac{\partial}{\partial z} E_x+ \frac{\partial}{\partial t} B_y, \hfill \\
\frac{\partial}{\partial y} cB_z =\frac{\partial}{\partial t} \frac{E_x}{c}+ \frac{\partial}{\partial z} c B_y. \hfill
\end{gathered}
\label{Eq. 1}
\end{equation}
For a wave oscillating with frequency $\omega$ and wavenumber $k_z=\omega/v_{ph}$ as $\propto \text{e}^{i(\omega t - k_z z)}$ 
the relation $\frac{1}{\partial t}=-\frac{v_{ph}}{\partial z}$ holds, thus
\begin{equation}
\begin{gathered}
 \frac{\partial }{{\partial x}}{ E_z} = \frac{\partial }{\partial z}\left( E_x-v_{ph}B_y \right), \hfill \\ 
-\frac{\partial }{{\partial y}}{cB_z} = \frac{\partial }{{\partial z}} \left(\frac{v_{ph}}{c}E_x-cB_y \right).\hfill
\end{gathered}
\label{Eq. 2}
\end{equation}
Equivalent transformations can be applied to the $E_y$ and $B_x$ components, resulting in
\begin{equation}
\begin{gathered}
 \frac{\partial }{{\partial y}}{ E_z} = \frac{\partial }{\partial z}\left( E_y+v_{ph}B_x \right),  \hfill \\
\frac{\partial }{{\partial x}}{cB_z} = \frac{\partial }{{\partial z}} \left(\frac{v_{ph}}{c}E_y+cB_x \right).\hfill
\end{gathered}
\label{Eq. 3}
\end{equation}
For a particle, traveling with longitudinal velocity component $v_z=v_{ph}$ synchronously with the wave, 
the first equations in Eqs~\ref{Eq. 2} and \ref{Eq. 3} mean:
\begin{equation}
\frac{1}{e} \frac{\partial}{\partial z} \vec F_{\perp}= -\frac{1}{e v_{ph}} \frac{\partial}{\partial t} \vec F_{\perp}= 
-i \frac{k_z}{e} \vec F_{\perp}= \vec{\nabla}_{\perp} E_z, 
\label{Eq. 4}
\end{equation}
where $\vec{\nabla}_{\perp} = \vec i_x \frac{\partial }{\partial x} +\vec i_y \frac{\partial }{\partial y}$, see~\cite{Garault} 
and references therein.\\
A synchronous transverse force $\vec F_t \neq 0$ can thus only exist together with a longitudinal component of 
the electric field $E_z$ in a deflecting rf field. 
The second equations in Eqs~\ref{Eq. 2} and \ref{Eq. 3} necessarily require the simultaneous existence of a longitudinal 
magnetic field component and for the particular case $v_z=v_{ph}=c$ the relation
\begin{equation}
c \left [ \vec i_z \times \vec{\nabla} B_z \right ]_{\perp}  = -i \frac{k_z}{e} \vec F_{\perp}= \vec{\nabla}_{\perp} E_z, 
\label{Eq. 5}
\end{equation}
holds.\\
A transverse force $\vec F_{\perp}$ is thus accompanied by both, a longitudinal electric, as well as a 
longitudinal magnetic field component.\\
A general cylindrical symmetric representation of the electro-magnetic fields in vacuum in a finite domain including the 
symmetry axis can be based on Hertzian basis vectors, see reference~\cite{Hahn} and references therein for a general discussion.
Table~\ref{Tab_1} summarizes the transverse TM--TE and the hybrid HM--HE basis vectors, as introduced 
in~\cite{Hahn}, with a modified notation (e.g. $\vec H$ has been replaced by $\vec B$). Here $k_0=\omega/c$, $k_z=\omega/v_{ph}$ and $k_r^2=k_0^2-k_z^2$ is 
used, while $n \in \mathbb{N}$  defines the azimuthal dependence.\\
\begin{table*}[ttt]   
\centering
\setlength{\tabcolsep}{8pt}
\caption{TM--TE and HM--HE basis vectors $\left( \displaystyle J_n=J_n(r k_r),\; J'_n=\frac{\partial J_n (k_r r)}{\partial (k_r r)}   
= \frac{n J_n(k_r r)}{k_r r} - J_{n+1}( k_r r) \right) $.}
\begin{tabular}{ l  c  c  c  c  c  c}
\hline
\hline
\\
 & TM  & TE  & HM   & HE & azimuthal & spatio-temporal  \\
& & &  &  & dependence  & dependence \\[2 mm]
\hline
\\
$\displaystyle E_r $         &  $\displaystyle  -k_z\frac{J'_n}{k_r^{n - 1}}$ & $\displaystyle  -\frac{n k_0}{r}\frac{J_n}{k_r^n}$ & $\displaystyle  k_0 k_z\frac{J_{n + 1}}{k_r^{n + 1}}$ & $\displaystyle   k_z^2\frac{J_{n + 1}}{k_r^{n + 1}}+\frac{n}{r}\frac{J_n}{k_r^n}$  &  $\displaystyle \cos(-n \vartheta)$ & $\displaystyle i\text{e}^{i(\omega t-k_z z)} $ \\[3 mm]

$\displaystyle E_\vartheta $ & $\displaystyle  \frac{n k_z}{r}\frac{J_n}{k_r^n}$ & $\displaystyle k_0\frac{J'_n}{k_r^{n - 1}}$    & $\displaystyle k_0 k_z\frac{J_{n + 1}}{k_r^{n + 1}}$ & $\displaystyle  k_0^2\frac{J_{n + 1}}{k_r^{n + 1}}- \frac{n}{r}\frac{J_n}{k_r^n} $ &  $\displaystyle \sin(n \vartheta)$ &	$\displaystyle i\text{e}^{i(\omega t-k_z z)}$ \\[3 mm]

$\displaystyle E_z         $ & $\displaystyle k_r^2\frac{J_n}{k_r^n}$         & 0                                                   & $\displaystyle k_0\frac{J_n}{k_r^n}$                 & $\displaystyle k_z\frac{J_n}{k_r^n}$                                                                & $\displaystyle \cos(-n \vartheta)$  & $ \displaystyle \text{e}^{i(\omega t-k_z z)}$ \\[3 mm]

$\displaystyle cB_r   $      & $\displaystyle  -\frac{n k_0}{r}\frac{J_n}{k_r^n}$ & $\displaystyle -k_z\frac{J'_n}{k_r^{n - 1}}$  & $\displaystyle  -k_z^2\frac{J_{n + 1}}{k_r^{n + 1}} -\frac{n}{r}\frac{J_n}{k_r^n} $ & $\displaystyle  -k_0 k_z\frac{J_{n + 1}}{k_r^{n + 1}}$ & $\displaystyle \sin(n \vartheta)$ & $\displaystyle i\text{e}^{i(\omega t-k_z z)}$   \\[3 mm]

$\displaystyle cB_\vartheta $ & $\displaystyle -k_0\frac{J'_n}{k_r^{n - 1}}$ & $\displaystyle  -\frac{n k_z}{r}\frac{J_n}{k_r^n}$ & $\displaystyle  k_0^2\frac{J_{n + 1}}{k_r^{n + 1}}-\frac{n}{r}\frac{J_n}{k_r^n} $ & $\displaystyle  k_0 k_z\frac{J_{n + 1}}{k_r^{n + 1}}$  & $\displaystyle  \cos(-n \vartheta) $ & $\displaystyle i\text{e}^{i(\omega t-k_z z)}$   \\[3 mm]

$\displaystyle cB_z  $        & 0                                             & $\displaystyle k_r^2\frac{J_n}{k_r^n}$              & $\displaystyle -k_z\frac{J_n}{k_r^n}$                & $\displaystyle -k_0\frac{J_n}{k_r^n}$                                                               & $\displaystyle  \sin(n \vartheta)$ & $\displaystyle \text{e}^{i(\omega t-k_z z)}$ \\[3 mm]
\hline
\hline        
\end{tabular}
\label{Tab_1}
\end{table*}
The expressions of the basis vectors in Table~\ref{Tab_1} are presented for $v_{ph} > c$ and $k_r^2 >0$. For waves with 
$v_{ph} < c$, $k_r$ becomes imaginary and the modified Bessel functions should be used to describe 
the radial dependencies of the field components.\\
In the domain under consideration any field component $G$ can be described by a linear combination 
of the basis vectors as:
\begin{equation}
G = A \times TM +B \times TE\quad  \text{or} \quad G= P\times HM+ Q\times HE
\label{Eq. 6}
\end{equation}
The coefficients $A$, $B$, $P$ and $Q$ are determined by the boundary conditions and the energy balance 
of the problem under consideration.\\
The traditional nomenclature of modes builds up on the TM--TE basis of transverse waves, which is well suited and usually 
applied for the field description in rf engineering and for modes with $n=0$ symmetry, as e.g. accelerating modes. 
TM (Transverse Magnetic) and TE (Transverse Electric) waves have each only one longitudinal field component; for TM waves $E_z \neq 0, B_z=0$ 
and for TE waves $E_z = 0, B_z \neq 0$. It is common practice to assume that the longitudinal field components 
and thus the coefficients $A$ and $B$ are independent; the discussion above shows however that this is not 
the general case for $n \ge 1$.\\
The general solution of the boundary conditions has thus six field components and requires the linear combination of two basis vectors, 
either TM and TE or HM and HE. Only for $n=0$ two separate solutions with three field components each can be formulated.\\ 
The TM-TE basis exhibits a methodical convergence problem when the phase velocity approaches $c$. 
For $v_{ph} \rightarrow c$, $k_z \rightarrow k_0$, $k_r \rightarrow 0$ the longitudinal components $E_z$ and $cB_z$ 
vanish as $k_r^2$. In addition all transverse field components vanish for $n=0$ but remain finite for $n \geq 1$, cf. Table~\ref{Tab_1}.\\
Comparing the longitudinal field components by means of Eq.~\ref{Eq. 6} yields the following relations of 
the coefficients of the two basis:
\begin{equation}
A = \frac{P k_0  + Q k_z}{k_r^2}, \quad
B = -\frac{P k_z  + Q k_0}{k_r^2}.
\label{Eq. 7}
\end{equation} 
$A$ and $B$ are thus divergent $\propto k_r^{-2}$ when $k_r$ approaches zero. Thus, while the basis vectors converge to zero, 
the product of basis vectors with the vector coefficients does not converge to zero and the field description is possible 
also in the limit $v_{ph}=c$.\\  
By means of the relations Eq.~\ref{Eq. 7} and the identity $ J'_n=\frac{n J_n}{r k_r} - J_{n + 1}$ the complete equivalence 
of the two basis can be shown.
The deflecting field representation for the transverse modes in a dielectric lined waveguide by Chang and Dawson~\cite{Dawson} 
in the TM--TE basis is hence essentially the same as the earlier result of Vagin and Kotov~\cite{Vagin} in 
the HM--HE basis.\\
The HM--HE basis has no convergence problem, i.e. no component of the basis converges to zero in the limit $k_r \rightarrow 0$ 
irrespective of $n$. Since each basis vector contains simultaneously electric and magnetic longitudinal 
vector components both, linear combinations and also pure HM or HE modes can satisfy the boundary conditions as well as 
Eq.~\ref{Eq. 4} and \ref{Eq. 5} for $n \ge 1$.\\
The description of pure TM or TE modes in the HM--HE basis leads to fixed relations of the vector coefficients as
$B=0$, $P k_z =-Q k_0$, $A= \frac{P}{k_0}$ for TM modes and $A=0$, $Pk_0 =-Qk_z $, $B= -\frac{Q}{k_0}$ for TE modes.
Still $A$ or $B$ need to be divergent in the limit $k_r \rightarrow 0 $ thus also $P$ and $Q$ get divergent while the vector 
components of the hybrid basis don't converge to zero. Thus only the sum of HM and HE remains finite in this case.\\
The TM--TE basis is therefore preferable for the field description in the case $n=0$, while the HM--HE basis is advantageous for $n \ge 1$.   
\section{Deflecting field for the relativistic case}
For the relativistic case $\beta_z =1, v_{ph}=c, k_0=k_z, k_r=0$ the Helmholtz wave equation reduces to the Laplace equation. 
In the HM--HE representation all basis vectors remain $\ne 0$ and continuous with respect to $k_r$. Expressions for the field 
components are found as limits for $k_r \rightarrow 0$ from the general form in Table~\ref{Tab_1} 
by expanding the Bessel function as:
\begin{equation}
\lim_{k_r \rightarrow 0} \frac{J_n(k_r r)}{k_r^n} = \frac{r^n}{2^n n!}.
\label{Eq. 7a}
\end{equation} 
The results of these transformations are summarized in Table~\ref{Tab_2}. Similar expressions can be found in \cite{Garault}, \cite{Hahn} and \cite{Vagin}.\\  
Following Table~\ref{Tab_2} the field distribution of the synchronous deflecting wave, $n=1, v_{ph}= c$ in the region of the 
interaction with the beam is:
\begin{equation}
\begin{gathered}
E_z=\left [ P+Q \right ] \frac{k_0 r}{2}\cos(\vartheta) \text{e}^{i(\omega t - k_0z)}, \hfill \\
E_r=i\left [ P \frac{k_0^2r^2}{8} + Q \left( \frac{k_0^2 r^2}{8}+\frac{1}{2} \right) \right ] \cos(\vartheta)\text{e}^{i(\omega t - k_0z)},\hfill \\
E_{\vartheta}= i\left [ P \frac{k_0^2r^2}{8} +Q \left( \frac{k_0^2r^2}{8}-\frac{1}{2} \right) \right ] \sin(\vartheta)\text{e}^{i(\omega t - k_0z)}, \hfill \\
cB_z=-\left [ P+Q \right ] \frac{k_0 r}{2} \sin(\vartheta) \text{e}^{i(\omega t - k_0z)},\hfill \\
cB_r=-i\left [ P \left(\frac{k_0^2r^2}{8}+\frac{1}{2} \right) + Q \frac{k_0^2 r^2}{8} \right ] \sin(\vartheta)  \text{e}^{i(\omega t - k_0z)},\hfill \\
cB_{\vartheta}=i\left [ P \left(\frac{k_0^2r^2}{8}-\frac{1}{2} \right)+Q \frac{k_0^2r^2}{8} \right ] \cos(\vartheta) \text{e}^{i(\omega t - k_0z)}, \hfill
\end{gathered}
\label{Eq. 8}
\end{equation}
Eq.~\ref{Eq. 8} describes for example the field in a dielectric lined waveguide~\cite{Vagin}.\\
For periodically iris loaded structures approximated expressions for the field components of the fundamental spatial harmonics 
inside the aperture $0 \leq r \leq a$ were obtained by means of the so-called small pitch approximation as (cf.~\cite{Garault}):  
\begin{equation}
\begin{gathered}
E_z = {\hat E} \frac{k_0 r}{2} \cos(\vartheta) \text{e}^{i(\omega t-k_0 z)},  \hfill \\
E_r=i {\hat E} \left[ \frac{k_0^2r^2+k_0^2 a^2}{8} \right]\cos(\vartheta) \text{e}^{i(\omega t-k_0 z)}, \hfill \\
E_{\vartheta}= i {\hat E} \left[ \frac{k_0^2r^2 - k_0^2 a^2}{8} \right]\sin(\vartheta)  \text{e}^{i(\omega t-k_0 z)}, \hfill\\
cB_z=-{\hat E} \frac{k_0 r}{2} \sin(\vartheta) \text{e}^{i(\omega t-k_0 z)}, \hfill \\  
cB_r=-i {\hat E} \left[\frac{4+k_0^2 r^2- k_0^2a^2}{8} \right]\sin(\vartheta) \text{e}^{i(\omega t-k_0 z)}, \hfill\\
cB_{\vartheta}= i {\hat E} \left[\frac{4-k_0^2r^2-k_0^2a^2}{8} \right]\cos(\vartheta) \text{e}^{i(\omega t-k_0 z)},  \hfill \\
\end{gathered}
\label{Eq. 9}
\end{equation}
The small pitch approximation requires that $v_{ph}=c$, that the cell length $d$ is shorter than the wavelength, $d \ll \lambda$, and 
that the iris thickness $t$ is smaller than the cell length, $t \ll d$, see Fig.~\ref{fig_1}, left. Furthermore it demands that the boundary 
condition $E_{\vartheta}=0$ is met at the aperture radius of the iris $r=a$.\\ 
A comparison of Eq.~\ref{Eq. 8} and \ref{Eq. 9} yields 
\begin{equation}
\begin{gathered}
P+Q={\hat E}, \quad P/Q=\frac{4}{k_0^2a^2}-1, \hfill \\
Q={\hat E} \frac{k_0^2 a^2}{4},\quad P={\hat E} \left ( 1- \frac{k_0^2 a^2}{4} \right ).
\end{gathered}
\label{Eq. 10}
\end{equation} 
The expression of the field components in Eq.~\ref{Eq. 9} are thus a particular case of the more general relations in 
Eq.~\ref{Eq. 8}, i.e. the expression for the deflecting field components in Eq.~\ref{Eq. 8} is valid for both, the wave in the 
longitudinally homogeneous dielectric lined waveguide and for the synchronous spatial harmonics in the 
periodically iris loaded structure.\\
The transverse force, Eq.~\ref{Eq. 0}, follows from Eq.~\ref{Eq. 8} for $v_z=c$ as:
\begin{equation}
\begin{gathered}
F_r= i e \frac{P+Q}{2} \cos(\vartheta) \text{e}^{i(\omega t - k_0 z)}= i e \frac{\hat E}{2} \cos(\vartheta) \text{e}^{i(\omega t - k_0 z)},\hfill \\
F_{\vartheta}=-i e \frac{P+Q}{2} \sin(\vartheta) \text{e}^{i(\omega t - k_0 z)}= -i e \frac{\hat E}{2} \sin (\vartheta) \text{e}^{i(\omega t - k_0 z)}, \hfill 
\end{gathered}
\label{Eq. 11}
\end{equation}
or, transferring to Cartesian coordinates, as: 
\begin{equation}
\begin{gathered}
F_x= i e \frac{\hat E}{2}  \text{e}^{ i(\omega t -k_0 z)}, \hfill \\
F_y=0, \hfill \\
E_z = \frac{\hat E}{2} k_0 x  \text{e}^{ i(\omega t -k_0 z)}, \hfill \\
cB_z= - \frac{\hat E}{2} k_0 y  \text{e}^{ i(\omega t -k_0 z)}. \hfill
\end{gathered}
\label{Eq. 12}
\end{equation} 
For the synchronous relativistic case $v_{ph}=v_z=c$ the deflecting force, or equivalently the deflecting field, is in the region of 
of the interaction $0 \leq r \leq a$ constant. 
The longitudinal field components are shifted by $\pi/2$ in the spatio-temporal phase and rise linearly with the distance from the axis.\\ 
According to Eq.~\ref{Eq. 10} is the deflecting field distributions in axially symmetric iris loaded structures for typical values of wave number 
$k_0$ and aperture radius $a$ dominated by the HM mode, because $P/Q > 1$. The ratio $P/Q $ depends on 
the design of the supporting structure and defines also other structure parameters, such as frequency, group velocity, effective shunt impedance and,
in case of periodically loaded structures, the level of higher spatial harmonics. 
Periodical structures with complexer geometry than the simple iris loaded structure give additional freedom for optimizations and 
allows to reach an overall more attractive set of parameters~\cite{FlPa2014}.
\begin{table*}[ttt]   
\centering
\setlength{\tabcolsep}{8pt}
\caption{Hybrid HM--HE solutions for $\displaystyle v_{ph}=c$.}
\begin{tabular}{ l  c  c  c  c}
\hline
\hline
\\
& HM   & HE   & azimuthal & spatio-temporal  \\
& & &  dependence  & dependence \\[2 mm]
\hline
\\
$\displaystyle E_r $	 & $\displaystyle  \frac{k_0^2 r^{n+1}}{2^{n+1}(n+1)!}$                                                  & $\displaystyle  \frac{k_0^2 r^{n + 1}}{2^{n + 1}(n+1)!}+\frac{r^{n-1}}{2^n (n-1)!} $ &  $\displaystyle \cos(-n\vartheta)$ & $\displaystyle i\text{e}^{i(\omega t-k_z z)} $ \\[3 mm]

$\displaystyle E_\vartheta $ & $\displaystyle  \frac{k_0^2 r^{n+1}}{2^{n+1}(n + 1)!}$                                                & $\displaystyle   \frac{k_0^2 r^{n + 1}}{2^{n+1} (n + 1)!}- \frac{r^{n-1}}{2^n (n-1)!} $ &  $\displaystyle \sin(n\vartheta)$ &	$\displaystyle i\text{e}^{i(\omega t-k_z z)}$ \\[3 mm]

$\displaystyle E_z         $ & $\displaystyle \frac{k_0 r^n}{2^n n!}$                                                                & $\displaystyle  \frac{k_0 r^n}{2^n n!}$                                                                & $\displaystyle \cos(-n\vartheta)$  & $ \displaystyle \text{e}^{i(\omega t-k_z z)}$ \\[3 mm]

$\displaystyle cB_r  $       & $\displaystyle -\frac{k_0^2 r^{n + 1}}{2^{n+1}(n + 1)!} -\frac{r^{n-1}}{2^n (n-1)!} $ & $\displaystyle  -\frac{k_0^2 r^{n + 1}}{2^{n + 1}(n+1)!}$                                              & $\displaystyle \sin( n\vartheta)$ & $\displaystyle i\text{e}^{i(\omega t-k_z z)}$   \\[3 mm]

$\displaystyle cB_\vartheta $ & $\displaystyle   \frac{k_0^2 r^{n + 1}}{2^{n + 1}(n+1)!}-\frac{r^{n-1}}{2^n (n-1)!} $ & $\displaystyle  \frac{k_0^2 r^{n + 1}}{2^{n + 1}(n+1)!}$                                               & $\displaystyle \cos(-n\vartheta) $ & $\displaystyle i\text{e}^{i(\omega t-k_z z)}$   \\[3 mm]

$\displaystyle cB_z  $       & $\displaystyle -\frac{k_0 r^n}{2^n n!}$                                                                & $\displaystyle  -\frac{k_0 r^n}{2^n n!}$                                                               & $\displaystyle  \sin(n\vartheta)$ & $\displaystyle \text{e}^{i(\omega t-k_z z)}$ \\[3 mm]
\hline
\hline        
\end{tabular}
\label{Tab_2}
\end{table*}
\section{Influence of the longitudinal field components on the particle dynamics}
>From Eq.~\ref{Eq. 4} follows that the integrated transverse momentum transfered to a particle moving on a straight line 
(rigid beam approximation) through a region with an arbitrary electromagnetic field is related to the integrated longitudinal 
field by   
\begin{equation}
p_{\perp}=-i \frac{e}{k_0 c}\int \limits_0^L \nabla_{\perp} E_z {dz}.
\label{Eq. 12a}
\end{equation}
Eq.~\ref{Eq. 12a} is referred to as Panofsky-Wenzel theorem. It is valid in this strict form only for synchronous motion 
with $v_z=v_{ph}=c$. (Asynchronous field components average to zero when the integration length is long enough.) The rigid 
beam approximation excludes ponderomotive forces, however, for a force like derived in Eq.~\ref{Eq. 12} (fundamental spatial 
harmonics, $n=1$, $v_{ph}=c$) ponderomotive forces are anyhow zero, because $F_y$ is everywhere zero -- not only one 
average -- and $F_x$ does not depend on $x$.\\
In accordance to Eq.~\ref{Eq. 12a} the transverse momentum $p_{\perp}=p_x$ and energy change of a bunch of particles passing 
through a deflecting structure of length $L$ following Eq.~\ref{Eq. 12} read as:
\begin{equation}
\begin{gathered}
p_x=\frac{eV}{c}\left(\sin(\varphi)+\cos(\varphi)\Delta \varphi \right) \hfill \\
E=e k_0 V x \left( \cos(\varphi)-\sin(\varphi)\Delta \varphi \right), \hfill \\
\end{gathered}
\label{Eq. 13}
\end{equation}
where $V=\frac{{\hat E}L}{2}$ is the integrated deflecting voltage and a first order Taylor expansion of the phase 
$\varphi=\omega t -k_0 z$ has been made. $\Delta \varphi$ denotes the position of a particle relative to the bunch 
center; $\Delta \varphi =-k_0 \Delta z$.\\ 
While the first term in the momentum equation describes the average momentum gained by the bunch, the second term describes 
the spread due to the differences experienced by particles in the head and the tail of the bunch. In the fully deflecting 
mode, $\varphi=\pi/2$, the momentum spread is to first order zero, while it is maximal at $\varphi=0$, the standard 
operation phase for cavity applications as diagnostics, crabbing and emittance exchange.\\
Due to the dependence of the longitudinal field on the transverse coordinate the induced energy spread is on all phases 
uncorrelated: 
\begin{equation}
\sigma_E= \;\left\{ \begin{gathered}
e k_0 V \sigma_x \quad \text{for} \quad \varphi=0, \hfill \\
e k_0^2 V \sigma_x \sigma_z \quad \text{for}\quad \varphi=\frac{\pi}{2}, \hfill \\ 
\end{gathered} \right.
\label{Eq. 14}
\end{equation}
with the transverse rms beam size in the streaking direction $\sigma_x$ and the longitudinal rms bunch length $\sigma_z$.
The Panofsky-Wenzel theorem, Eq.~\ref{Eq. 12a}, as well as Eq.~\ref{Eq. 13} and Eq.~\ref{Eq. 14} describe the beam dynamics 
to first order. In second order the induced transverse momentum couples to the longitudinal magnetic field and the 
transverse particle position changes, which leads to an additional correlated energy spread of~\cite{Emma 2011}
\begin{equation}
\frac{{\Delta {E^{cor}}}}{{\Delta z}} = \frac{{\left( {ek_0V} \right)}^2}{c p_z}\frac{L}{6},
\label{Eq. 15}
\end{equation}
where $p_z$ denotes the longitudinal momentum of the particle.\\
The second order effects combine the cosine-like transverse momentum with the sine-like longitudinal field components, the 
Taylor expansion of the product is thus of the form $\frac{1}{2} \sin(2\varphi)+\cos(2\varphi)\Delta \varphi$ and 
Eq.~\ref{Eq. 15} is therefore valid for $\varphi=0$ and for $\varphi=\pi/2$.\\
The momentum in the streaking direction leads in combination with the longitudinal magnetic field also to a force in 
direction perpendicular to the streaking plane:
\begin{equation}
{\hat F_y}=e\frac{p_x}{\gamma m_0}B_z,
\label{Eq. 16}
\end{equation}
with the rest mass of the particle $m_0$ and the relativistic Lorentz factor $\gamma$. Thus
\begin{equation}
p_y = \frac{1}{c}\int {{\hat F_y}} \;dz =\frac{{\left( {ek_0V} \right)}^2}{2c^2 p_z} y \Delta z.
\label{Eq. 17}
\end{equation}
The transverse momentum $p_y$ is linear in the transverse position $y$, i.e. it is a focusing force, which depends 
however on the longitudinal position $\Delta z$ in the bunch. It leads thus to a projected emittance contribution of
\begin{equation}
\varepsilon_y=\frac{{\left( {ek_0V} \right)}^2}{2m_0c^2 p_z} \sigma_y^2 \sigma_z,
\label{Eq. 18}
\end{equation}
with the transverse rms beam size perpendicular to the streaking direction $\sigma_y$.
Eq.~\ref{Eq. 17} and \ref{Eq. 18} are again valid for $\varphi=0$ and for $\varphi=\pi/2$.\\    
The emittance growth Eq.~\ref{Eq. 18} in the direction perpendicular to the streaking force is a fundamental property 
of the deflecting field which doesn't depend on the design of the supporting structure or the operating mode of the cavity.\\
For present day beam and cavity parameters the emittance growth is very small, but it can not be eliminated.
Additional transverse forces can still appear due the spatial harmonics especially in the end cells of rf structures 
and due to the backward traveling component in standing wave cavities. These forces average out in a first order 
approximation of the particle motion, but are relevant for second order effects~\cite{Rei 97, Opanasenko} and often dominate the beam dynamics in the plane perpendicular to the streaking direction. 
\section{Summary}
Transverse deflecting rf structures find nowadays numerous applications in particle accelerators. Regardless of the 
design and operating mode of the supporting structure, a synchronous transverse force, generated by the common interaction of the 
transverse electric and magnetic field components, is always accompanied by both, electric 
as well as magnetic, longitudinal field components. 
The complete deflecting rf field has necessarily six field components and requires the representation  
by a linear combination of two basis vectors. The description in  the HM--HE basis avoids convergence problems 
which are characteristic for the usual TM--TE basis for waves matched to the velocity of light.\\
Both, the longitudinal electric and the longitudinal magnetic field component lead 
to undesired and fundamental beam dynamics effects.\\ 
The longitudinal magnetic field, which has been ignored so far, in combination with the induced transverse momentum in the 
direction of the deflecting force, results in a small, but fundamental, emittance contribution in the direction perpendicular to the deflecting 
force. 
\section{Acknowledgments} 
The author VP is supported by RFMEFI62117X0014 program.    

\end{document}